\documentclass[a4paper]{jpconf}
\usepackage{graphicx}
\begin{document}
\title{The Capabilities of Monochromatic EC Neutrino Beams with the SPS Upgrade}

\author{Catalina Espinoza and Jos\'e Bernab\'eu}

\address{IFIC, Universidad de Valencia-CSIC, E-46100, Burjassot, Valencia, Spain }

\ead{m.catalina.espinoza@uv.es, jose.bernabeu@uv.es}

\begin{abstract}
The goal for future neutrino facilities is the determination of the U(e3) mixing and CP violation in neutrino oscillations. This will require precision experiments with a very intense neutrino source and energy control. With this objective in mind, the creation of monochromatic neutrino beams from the electron capture decay of boosted ions by the SPS of CERN has been proposed. We discuss the capabilities of such a facility as a function of the energy of the boost and the baseline for the detector. We conclude that the SPS upgrade to 1000 GeV is crucial to reach a better sensitivity to CP violation iff it is accompanied by a longer baseline. We compare the physics potential for two different configurations: I) $\gamma=90$ and $\gamma=195$ (maximum achievable at present SPS) to Frejus; II) $\gamma=195$ and $\gamma=440$ (maximum achievable at upgraded SPS) to Canfranc. The main conclusion is that, whereas the gain in the determination of U(e3) is rather modest, setup II provides much better sensitivity to CP violation.
\end{abstract}

\section{Introduction}

A number of experimental facilities to significantly improve on present
sensitivity to the connecting mixing $\theta_{13}$ in neutrino oscillations and have access to the CP-violating phase $\delta$ have been discussed in the literature: neutrino factories (neutrino beams from boosted-muon 
decays), superbeams (very intense conventional neutrino beams), improved reactor experiments and more recently $\beta$-beams \cite{zucchelli}.
The original standard scenario for beta beams with lower $\gamma=60/100$ and  short baseline $L=130$~Km from CERN to Frejus with $^6He$ and $^{18}Ne$ ions has seen a variant by using an electron capture facility for monochromatic neutrino beams \cite{Bernabeu:2005jh}. New proposals also include the high $Q$ value $^8Li$ and $^8Be$ isotopes in a $\gamma=100$ facility \cite{Rubbia:2006pi}. In this paper we discuss the physics reach that a high energy facility for  EC beams may provide with the expected SPS upgrade at CERN. Such study has been made for the standard beta beam facility \cite{Burguet-Castell:2005pa}.
In Section~2 we discuss the virtues of the suppressed oscillation channel $(\nu_e \to \nu_\mu)$ in order to have access to the parameters $\theta_{13}$ and $\delta$. The interest of energy dependence, as obtainable in the EC facility, is emphasized.  In Section~3 we present new results on the comparison between (low energies, short baseline) and (high energies, long baseline) configurations for an EC facility with a single ion. Section~4 gives our conclusions.

\section{Suppressed Neutrino Oscillation}

The observation of $CP$ violation needs an experiment in which the emergence of another neutrino flavour is detected rather than the deficiency of the original flavour of the neutrinos. At the same time, the interference needed to generate CP-violating observables can be enhanced if both the atmospheric and solar components have a similar magnitude. This happens in the suppressed $\nu_e \to \nu_{\mu}$ transition. The appearance probability $P(\nu_e \to \nu_{\mu})$ as a function of the distance between source and detector $(L)$ is given by \cite{cervera}

\begin{eqnarray}\label{prob}
P({\nu_e \rightarrow \nu_\mu})  \simeq ~ 
s_{23}^2 \, \sin^2 2 \theta_{13} \, \sin^2 \left ( \frac{\Delta m^2_{13} \, L}{4E} \right ) + ~   c_{23}^2 \, \sin^2 2 \theta_{12} \, \sin^2 \left( \frac{ \Delta m^2_{12} \, L}{4E} \right ) 
\nonumber \\
 + ~ \tilde J \, \cos \left ( \delta - \frac{ \Delta m^2_{13} \, L}{4E} \right ) \;
\frac{ \Delta m^2_{12} \, L}{4E} \sin \left ( \frac{  \Delta m^2_{13} \, L}{4E} \right ) \,,
\end{eqnarray}
where $\tilde J \equiv c_{13} \, \sin 2 \theta_{12} \sin 2 \theta_{23} \sin 2 \theta_{13}$. The three terms of Eq.~(\ref{prob}) correspond, respectively, to contributions from the
atmospheric and solar sectors and their interference. As seen, the $CP$ violating contribution
has to include all mixings and neutrino mass differences to become observable. The four measured parameters $(\Delta m_{12}^2,\theta_{12})$  and  $(\Delta m_{23}^2,\theta_{23})$ have been fixed throughout this paper to their mean values \cite{Gonzalez-Garcia:2004jd}. 

From general arguments of CPT invariance and absence of absorptive parts the CP-odd probability is odd in time and then odd in the baseline L (formally). Vacuum oscillations are only a function of $E/L$ so that, at fixed $L$, the CP-odd probability is odd in the energy (formally). This proves that the study of neutrino oscillations in terms of neutrino energy will be able to separate out the CP phase $\delta$ from the mixing parameters. A control of this energy may be obtained from the choice of the boost in the EC facility with a single ion. In the beams with a continuous spectrum, the neutrino energy has to be reconstructed in the detector. In water-Cerenkov detectors, this reconstruction is made from supposed quasielastic events by measuring both the energy and direction of the charged lepton. This procedure suffers from non-quasielastic background, from kinematic deviations due to the nuclear Fermi momentum and from dynamical suppression due to exclusion effects \cite{Bernabeu72}. For EC decay,
the rare-earth nuclei above $^{146}Gd$ have a small enough half-life for
 electron capture processes. This is in contrast with the proposal of EC beams  with fully stripped long-lived ions \cite{Sato:2005ma}.  We discuss the option of short-lived ions \cite{Bernabeu:2005jh}.
\begin{figure}[h]
\includegraphics[width=13pc, angle=-90]{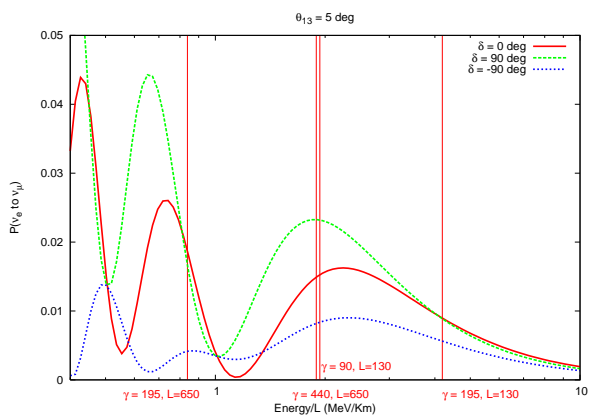}\hspace{2pc}%
\begin{minipage}[t]{14pc}\caption{\label{proba}The appearance pro\-ba\-bi\-li\-ty $P(\nu_e \rightarrow \nu_{\mu})$ for neutrino oscillations
 as a function of E/L, with fixed  connecting
mixing. The three curves refer to different values of the CP violating phase $\delta$. The vertical lines are the energies of our simulation study in the EC facility.}
\end{minipage}
\end{figure}
\section{EC-Beam Capabilities at Different Energies and Baselines}

The neutrinos emerging from a boosted ion beam decaying by EC are concentrated inside a narrow cone around the forward direction. If the ions
 are kept in the decay ring longer than the half-life, the energy distribution of the Neutrino 
Flux arriving to the detector in absence of neutrino oscillations is given by the Master Formula

\begin{eqnarray}\label{master}
\frac{d^2N_\nu}{ dS dE}
=  \frac{1}{\Gamma} \frac{d^2\Gamma_\nu}{dS dE} N_{ions} \simeq  \frac{\Gamma_\nu}{\Gamma} \frac{ N_{ions}}{\pi L^2} \gamma^2
\delta{\left(E - 2 \gamma E_0 \right)},
\end{eqnarray}
with a dilation factor $\gamma >> 1$. It is remarkable that the result is given only in terms of the branching ratio 
and the neutrino energy and independent of nuclear models. In Eq.~(\ref{master}), $N_{ions}$ is the total number of 
 ions decaying to neutrinos. As a result, such a facility will measure the neutrino oscillation parameters by changing the $\gamma$'s of the decay ring (energy dependent measurement) and there is no need of energy reconstruction in the detector.

For the study of the physics reach associated with such a facility, we combine two different energies for the same $^{150}Dy$ ion using two Setups. In all cases we consider $10^{18}$ decaying ions/year, a water Cerenkov Detector with fiducial mass of $440$~Kton and both appearance ($\nu_{\mu}$) and disappearance ($\nu_e$) events. Setup I  is associated with a five year run at $\gamma=90$ (close to the minimum energy to avoid atmospheric neutrino background) plus a five year run at $\gamma=195$ (the maximum energy achievable at present SPS), with a baseline $L=130$~Km from CERN to Frejus. The results for Setup I are going to be compared with those for Setup II, associated with a five year run at $\gamma=195$ plus a five year run at $\gamma=440$ (the maximum achievable at the upgraded SPS with Proton energy of $1000$~GeV), with a baseline $L=650$~Km from CERN to Canfranc.

\begin{figure}[h]
\begin{minipage}{16pc}
\includegraphics[width=12pc, angle=-90]{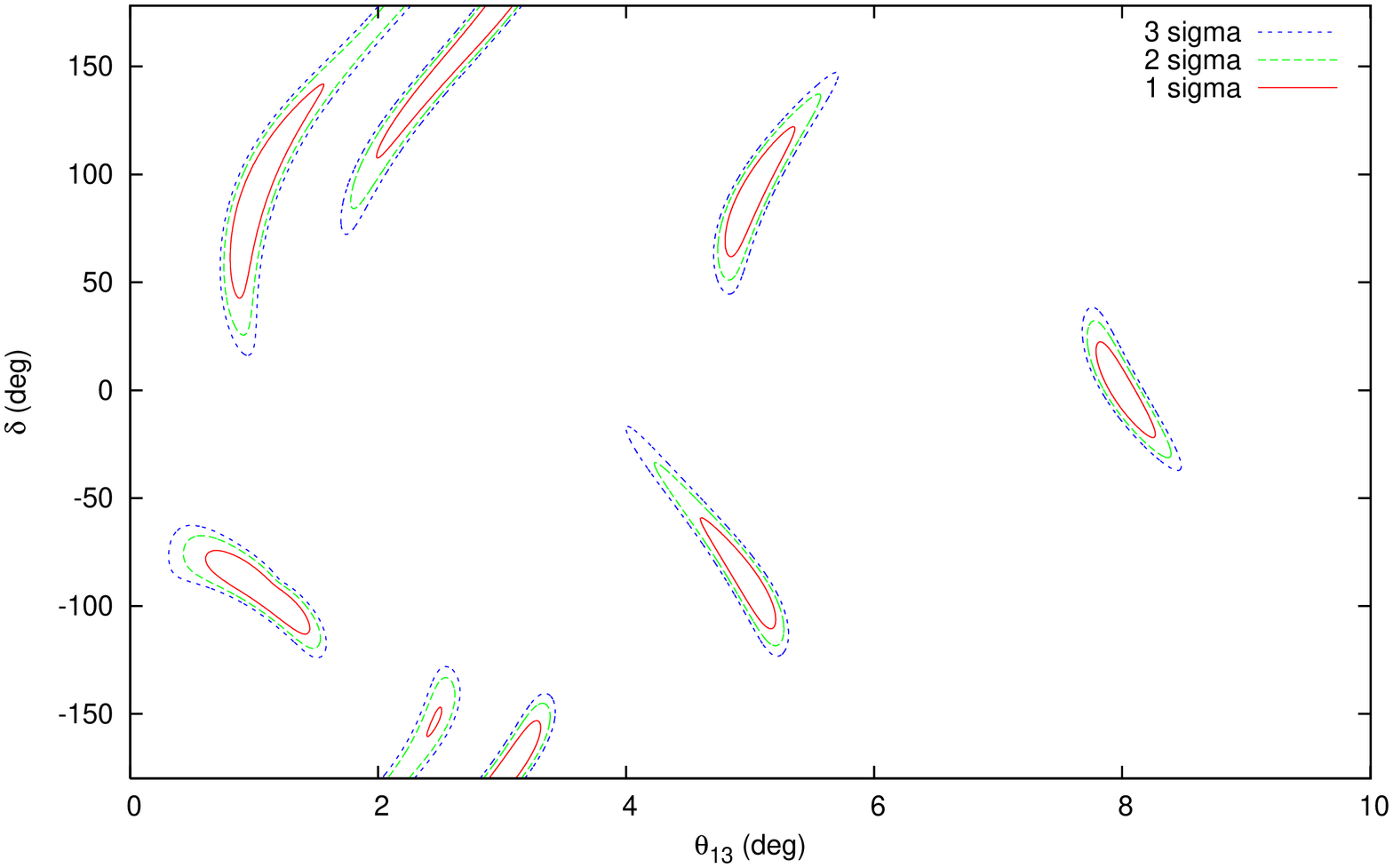}
\vspace*{-2pc}\caption{\label{fit-setupI}Setup I. Fit for $(\theta_{13}, \delta)$ from statistical distribution.}
\end{minipage}\hspace{3pc}%
\begin{minipage}{16pc}
\includegraphics[width=12pc, angle=-90]{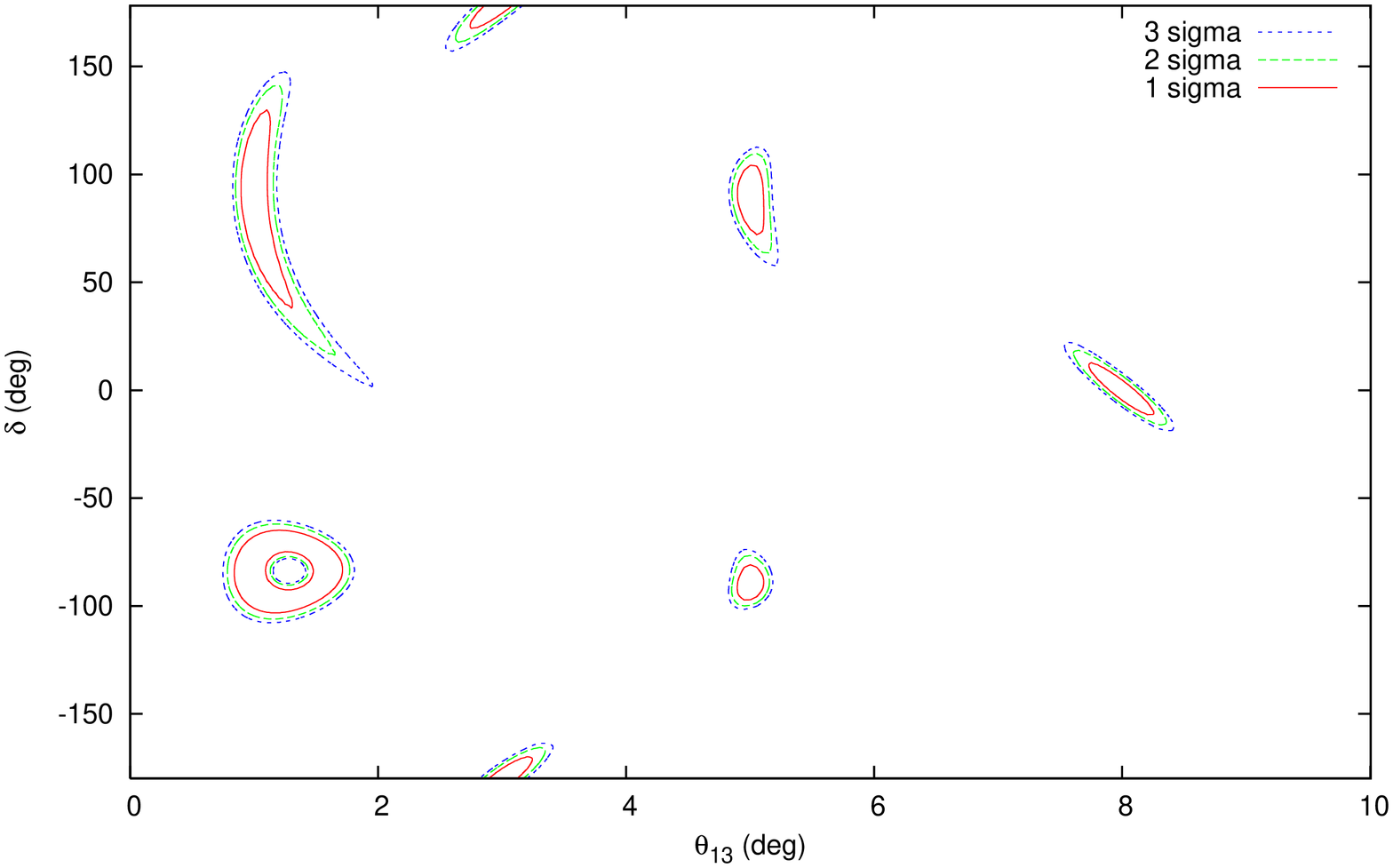}
\vspace*{-2pc}\caption{\label{fit-EC}Setup II. Fit for $(\theta_{13}, \delta)$ from statistical distribution.}
\end{minipage} 
\end{figure}

\begin{figure}[h]
\begin{minipage}{16pc}
\includegraphics[width=12pc, angle=-90]{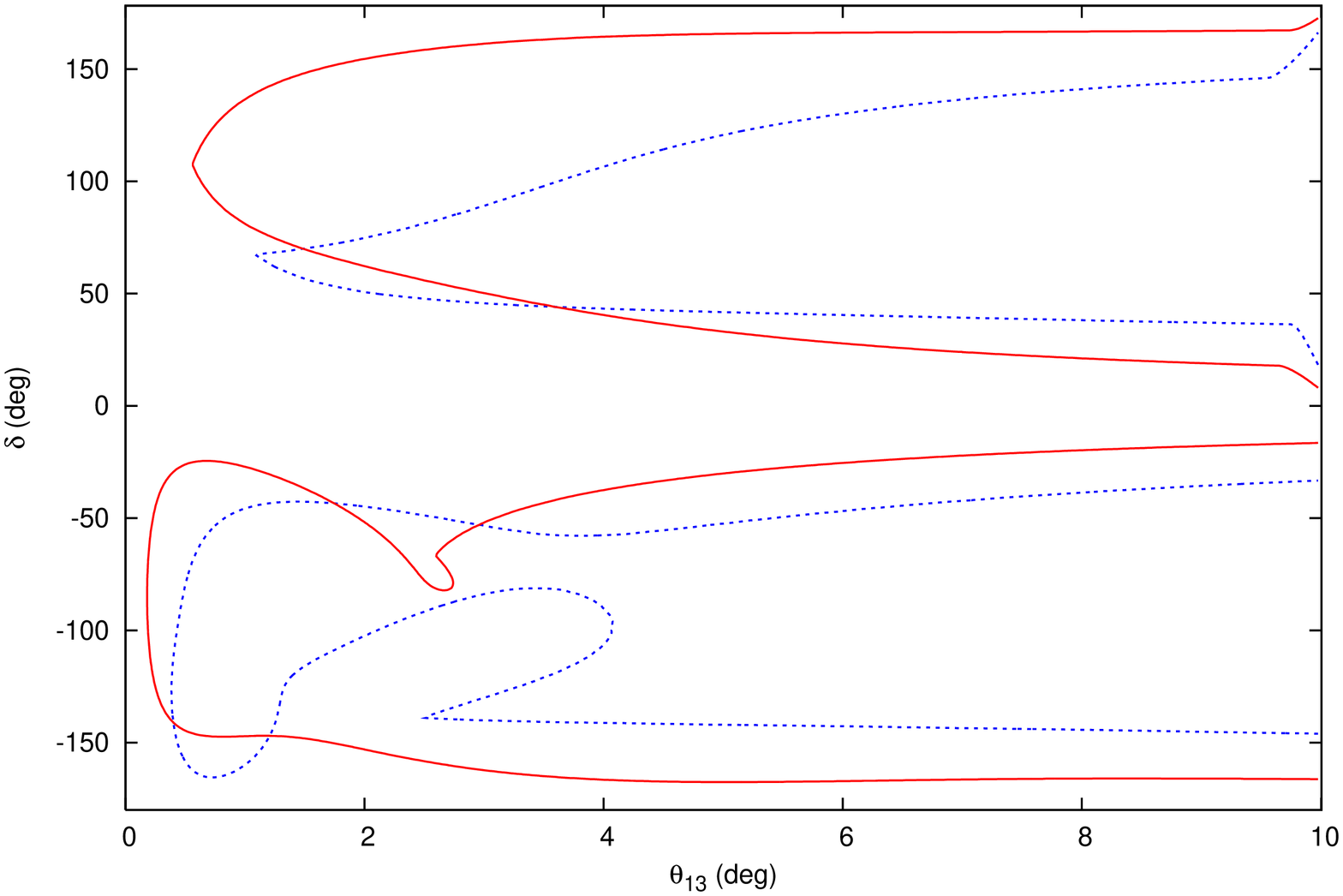}
\vspace*{-1pc}\caption{\label{fit-EC-Des}CP Violation Exclusion plot at 99$\%$ CL, if $\theta_{13}$ is still unknown, for the two reference Setups: I (broken blue line) and II (continuous red line).}
\end{minipage}\hspace{3pc}%
\begin{minipage}{16pc}
\includegraphics[width=12pc, angle=-90]{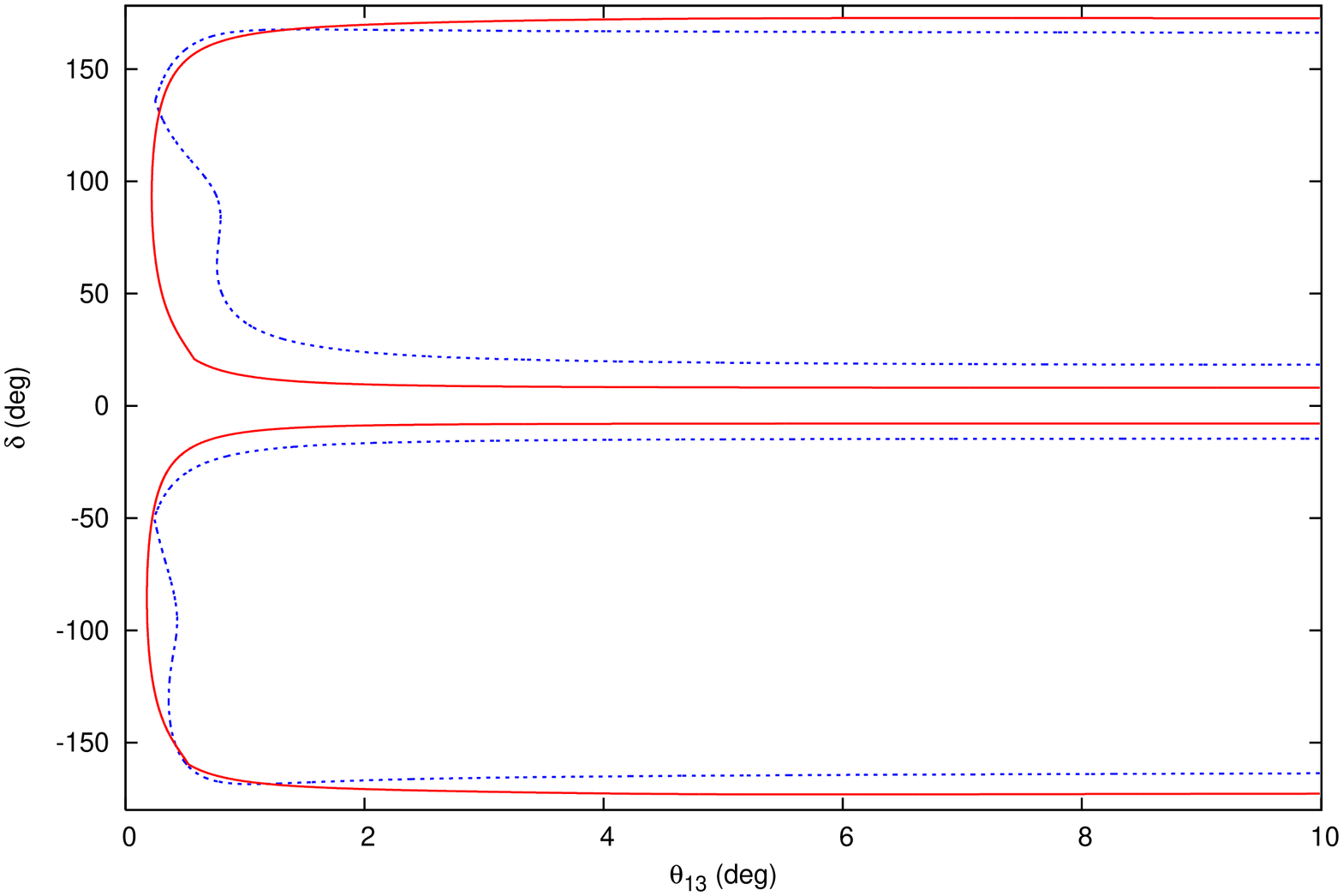}
\vspace*{-1pc}\caption{\label{fit-EC-Con}CP Violation Exclusion plot at 99$\%$ CL, if $\theta_{13}$ is previously known, for the two reference Setups: I (broken blue line) and II (continuous red line).}
\end{minipage} 
\end{figure}

For the Setup I we generate the statistical distribution of events from  assumed values of $\theta_{13}$ and $\delta$. The corresponding fit is shown in Fig.~\ref{fit-setupI} for selected values of $\theta_{13}$ from $8^o$ to $1^o$ and covering a few values of the CP phase $\delta$. As observed, the principle of an energy dependent measurement (illustrated here with two energies) is working to separate out the two parameters. With this configuration the precision obtainable for the mixing (even at 1 degree) is much better than that for the CP phase. The corresponding exclusion plot which defines the sensitivity to discover  CP violation $\delta \ne 0, 180^o$ is presented in Fig.~\ref{fit-EC-Des}. For $99\%$~CL the sensitivity to see CP violation becomes significant for $\theta_{13}>4^o$.

In the case of Setup II the longer baseline for $\gamma=195$ leads to a value of $E/L$ well inside the second oscillation (see Fig.~\ref{proba}). In that case the associated strip in the ($\theta_{13}$, $\delta$) plane has a more pronounced curvature, so that the two parameters can be better disantangled. The statistical distribution generated for some assumed values of ($\theta_{13}$, $\delta$) has been fitted and the $\chi^2$ values obtained. The results are given in Fig.~\ref{fit-EC}. Qualitatively, one notices that the precision reachable for the CP phase is much better than that for Setup I. One should emphasize that this improvement in the CP phase has been obtained with the neutrino channel only, using two appropriate different energies. The corresponding exclusion plot which defines the sensitivity to discover  CP violation $\delta \ne 0, 180^o$ is presented in Fig.~\ref{fit-EC-Des}. For $99\%$~CL the sensitivity  to see CP violation becomes now  significant even for values  of $\theta_{13}$ smaller than those for Setup I.

At the time of the operation of this proposed Facility  it could happen that the connecting mixing $\theta_{13}$ is already known from the approved experiments for second generation neutrino oscillations, like Double CHOOZ, T2K and NOVA. To illustrate the gain obtainable in the sensitivity to discover CP violation from the previous knowledge of $\theta_{13}$ we present in Fig.~\ref{fit-EC-Con} the expected sensitivity with the  distribution of events depending on a single parameter $\delta$ for a fixed known value of $\theta_{13}$. The result is impressive: even for a mixing angle of one  degree, the CP violation sensitivity  at $99\%$~CL reaches values around $10^o$ for Setup II.

\section{Conclusions}
The simulations of the physics output for EC beams indicate: 1) The upgrade to higher energy ($E_p=1000$~GeV) is crucial to have a better sensitivity to CP violation, which is the main objective of the next generation neutrino oscillation experiments, iff accompanied by a longer baseline; 2) The best $E/L$ in order to have a  higher sensitivity to the mixing $U(e3)$ is not the same than that for the CP phase. Like the phase-shifts, the presence of $\delta$ is easier to observe when the energy of the neutrino beams enters  into the region of the second oscillation. The mixing is better seen around the first oscillation maximum, instead. In particular, Setup II in EC beams, i.e., with $\gamma's$ between $195$ and $440$ and a baseline $L=650$~Km (Canfranc),  has an impressive sensitivity to CP violation, reaching a sensitivity around $20^o$, for $99\%$~CL, or better (if some knowledge on the value of $\theta_{13}$ is previously established).  

Besides the feasibility studies for the machine, most important for physics is the study of the optimal configuration by combining low energy with high energy neutrino beams, short baseline with long baseline and/or EC monochromatic neutrinos with $^6He$ $\beta^-$ antineutrinos.

\end{document}